\documentstyle[11pt]{article}
\def\PRL #1 #2 #3{{\sl Phys.  Rev.  Lett.} {\bf#1} (#2) #3}
\def\NPB #1 #2 #3{{\sl Nucl.  Phys.} {\bf B #1} (#2) #3}
\def\NPBFS #1 #2 #3 #4{{\sl Nucl.  Phys.} {\bf B #2} [FS#1] (#3) #4}
\def\CMP #1 #2 #3{{\sl Commun.  Math.  Phys.} {\bf #1} (#2) #3}
\def\PRD #1 #2 #3{{\sl Phys.  Rev.} {\bf D #1} (#2) #3}
\def\PLA #1 #2 #3{{\sl Phys.  Lett.} {\bf #1A} (#2) #3}
\def\PLB #1 #2 #3{{\sl Phys.  Lett.} {\bf B #1} (#2) #3}
\def\JMP #1 #2 #3{{\sl J.  Math.  Phys.} {\bf #1} (#2) #3}
\def\PTP #1 #2 #3{{\sl Prog.  Theor.  Phys.} {\bf #1} (#2) #3}
\def\SPTP #1 #2 #3{{\sl Suppl.  Prog.  Theor.  Phys.} {\bf #1} (#2) #3}
\def\AoP #1 #2 #3{{\sl Ann.  of Phys.} {\bf #1} (#2) #3}
\def\PNAS #1 #2 #3{{\sl Proc.  Natl.  Acad.  Sci.  USA} {\bf #1} (#2) #3}
\def\RMP #1 #2 #3{{\sl Rev.  Mod.  Phys.} {\bf #1} (#2) #3}
\def\PR #1 #2 #3{{\sl Phys.  Reports} {\bf #1} (#2) #3}
\def\AoM #1 #2 #3{{\sl Ann.  of Math.} {\bf #1} (#2) #3}
\def\UMN #1 #2 #3{{\sl Usp.  Mat.  Nauk} {\bf #1} (#2) #3}
\def\FAP #1 #2 #3{{\sl Funkt.  Anal.  Prilozheniya} {\bf #1} (#2) #3}
\def\FAaIA #1 #2 #3{{\sl Functional Analysis and Its Application}
{\bf #1} (#2) #3}\def\BAMS #1 #2 #3{{\sl Bull.  Am.  Math.  Soc.} {\bf #1} (#2) #3} 
\def\TAMS #1 #2 #3{{\sl Trans.  Am.  Math.  Soc.} {\bf #1} (#2) #3}
\def\InvM #1 #2 #3{{\sl Invent.  Math.} {\bf #1} (#2) #3}
\def\LMP #1 #2 #3{{\sl Letters in Math.  Phys.} {\bf #1} (#2) #3}
\def\IJMPA #1 #2 #3{{\sl Int.  J.  Mod.  Phys.} {\bf A #1} (#2) #3}
\def\AdM #1 #2 #3{{\sl Advances in Math.} {\bf #1} (#2) #3}
\def\RMaP #1 #2 #3{{\sl Reports on Math.  Phys.} {\bf #1} (#2) #3}
\def\IJM #1 #2 #3{{\sl Ill.  J.  Math.} {\bf #1} (#2) #3}
\def\APP #1 #2 #3{{\sl Acta Phys.  Polon.} {\bf #1} (#2) #3}
\def\TMP #1 #2 #3{{\sl Theor.  Mat.  Phys.} {\bf #1} (#2) #3}
\def\JPA #1 #2 #3{{\sl J.  Physics} {\bf A#1} (#2) #3}
\def\JSM #1 #2 #3{{\sl J.  Soviet Math.} {\bf #1} (#2) #3}
\def\MPLA #1 #2 #3{{\sl Mod.  Phys.  Lett.} {\bf A #1} (#2) #3}
\def\JETP #1 #2 #3{{\sl Sov.  Phys.  JETP} {\bf #1} (#2) #3}
\def\JETPL #1 #2 #3{{\sl Sov.  Phys.  JETP Lett.} {\bf #1} (#2) #3}
\def\PHSA #1 #2 #3{{\sl Physica} {\bf A #1} (#2) #3}
\def\CQG #1 #2 #3{{\sl Class.  Quantum Grav.} {\bf #1} (#2) #3}
\def\SJNP #1 #2 #3{{\sl Sov. J.  Nucl. Phys. (Yadern.Fiz.)} {\bf #1} (#2) #3}
\def\d{\delta}\def\e{\epsilon}

\def\s{\sigma}
\def\Th{\Theta}

\newcommand{\p}[1]{(\ref{#1})}
\newcommand{\plabel}{\label}
\begin{document}
\renewcommand{\thefootnote}{\fnsymbol{footnote}}
\setcounter{page}0
\thispagestyle{empty}
\begin{flushright}
{\bf 
hep-th/9905144 \\
TUW/99--09 \\
1999, May 19} \\
\end{flushright}

\begin{center}
{\LARGE 
Current Density Distributions and \\ 
a Supersymmetric Action for Interacting Brane Systems} 

\vspace{1.0cm}
\renewcommand{\thefootnote}{\dagger} \vspace{0.2cm}

\bigskip 

{\bf Igor  Bandos}\footnote{Lise Meitner Fellow. Also at the 
{\it   Institute for Theoretical Physics, 
NSC Kharkov Institute of Physics and Technology, 
 310108, Kharkov,  Ukraine 
 e-mail: bandos@kipt.kharkov.ua}} 
 and  {\bf Wolfgang Kummer} 

\bigskip 

{\it  Institut f\"{u}r Theoretische Physik, 
\\ 
Technische
 Universit\"{a}t Wien, 
\\ Wiedner Hauptstrasse 8-10, A-1040 Wien\\
 e-mail: wkummer@tph.tuwien.ac.at \\
bandos@tph32.tuwien.ac.at \\
}

\vspace{0.5cm}

{\bf Abstract}
 \end{center}

\medskip 

{\small 

We propose a method to obtain a manifestly supersymmetric action functional 
for interacting brane systems. It is based on the induced map 
of the worldvolume of low-dimensional branes into the  worldvolume of 
the space-time filling brane ((D-1)-brane), which may be 
either dynamical or auxiliary,  
and implies an identification of Grassmann coordinate fields
of lower dimensional branes 
with an image of the Grassmann coordinate fields of that (D-1)-brane. 
With this  identification the covariant current distribution forms 
with  support on the superbrane worldvolumes become invariant under the 
target space supersymmetry 
and can be used to write the coupled superbrane action as an integral over the 
D-dimensional manifolds ((D-1)-brane worldvolume).  
We compare the equations derived from  this new 
('Goldstone fermion embedded') action   
with the ones produced by a more straightforward generalization 
of the free brane actions 
 based on the incorporation of the boundary terms with Lagrange 
multipliers ('superspace embedded' action). 
We find that both procedures produce the same equations of motion 
and thus  justify each other. 
Both actions are presented explicitly for the coupled system of 
a $D=10$ super-D3-brane and a 
fundamental 
superstring which ends on the super-D3-brane.}

\medskip 

\setcounter{page}1
\renewcommand{\thefootnote}{\arabic{footnote}} \setcounter{footnote}0
\setcounter{page}0

\newpage

\bigskip

\section*{Introduction}

Intensive studies of interacting 
branes  (intersecting branes and branes ending on branes) 
\cite{inters1}--\cite{West} 
were performed 
for the pure bosonic limit \cite{SUGRA,inters1,Sato} 
or in the framework of the 
'probe brane' approach \cite{Mourad,Sezgin25}.  
In spite of many successes achieved in this way, it is desirable to obtain a 
 complete and manifestly supersymmetric description of 
the interacting brane systems 
at  the level of (quasi)classical effective action.

Actually, the preservation of symmetries 
in the presence of boundaries (including the boundaries of open branes 
ending on other branes)  
requires a consideration of anomalies \cite{HW,Mourad}, while at the  
 classical level  the boundary 
breaks  at least half of the supersymmetry \cite{gsw,typeI,Mourad} 
(see Appendix A). 
So at that level
one may 
search for 
an action for a coupled brane system, which includes manifestly supersymmetric 
bulk terms for all the branes and allows direct variations.

In this paper we propose a procedure
to obtain such a supersymmetric  
action for an interacting 
 brane system. 
It involves 
a (dynamical or auxiliary) space-time 
filling brane ('(D-1)-brane dominance') and uses the identification of 
all the Grassmann coordinate fields  of lower dimensional branes 
$\hat{\Theta} (\xi )$ with 
images of the (D-1)-brane Grassmann coordinate fields 
$\hat{\Theta} (\xi ) = \Theta (x(\xi))$. 
In this sense intersecting branes are considered as 
embedded into the Goldstone fermion theory \cite{VA}
(which is just the space-time 
filling superbrane \cite{GalperinBagger})
rather then 
into superspace.   
As a result we obtain less fermionic equations 
than expected. 
The equations can be split in separate ones for 
the open brane and the host brane, but with an {\sl indefinite} 
source localized at the intersection.

To justify the above result  we turn to a 
more straightforward generalization 
of the free superbrane actions  to the coupled brane system.  
It  produces the necessary identification 
of the supercoordinate function at the intersection 
by an incorporation of a bosonic vector 1-form and a Grassmann spinor 
1-form Lagrange multipliers into the boundary term. 
(To our knowledge 
such a quite simple action was not considered before). 
We find that the Lagrange multipliers involve an ambiguity in the 
equations of motion. Due to that ambiguity 
the equations of motion obtained from 
those ('superspace embedded' or SSPE)
 action 
 and ones derived from the above mentioned 
 ('Goldstone fermion embedded' or GFE) 
action functional turns out to be  equivalent. 
Thus the two approaches justify  each other.

We find that an ambiguity in the sources localized at the intersection 
appears in the bosonic coordinate equations as well and, thus, have to be taken into account even in the pure bosonic limit of the coupled brane system.  

We present an explicit form of 
 both SSPE and GFE actions  
for the system of the closed super-D3-brane and the fundamental 
superstring ending on the super-D3-brane 
\footnote{
This system is a special one for string perturbation theory. 
Here D-branes are considered as  
(sub)manifolds where the open string endpoints live upon 
and can be described by open string states 
(see \cite{typeI} and refs. therein).  
However, for the description in the language of  brane effective action 
functionals, which is considered here, 
this system provides a quite  generic example
of interacting branes 
(brane democracy \cite{T95}). }. 
The latter is of  particular interest for String/M-theory \cite{DbraneM}, 
its applications to gauge theory 
\cite{witten96},  
as well as  in the frame of the Maldacena conjecture 
\cite{Maldacena}. 
To be concrete, 
we describe our approach just for 
this specific system.

\section{An action with Lagrange multipliers}

The actions of a free type $IIB$ superstring and a free 
super-D3-brane can be presented as integrals of a Lagrangian 
2--form 
${\cal L}_2^{IIB}$ 
and 4-form 
${\cal L}_4^{D3}$ 
\begin{equation}\plabel{S0IIB}
S^{IIB}_{0}= \int_{{\cal M}^{1+1}}  {\cal L}^{IIB}_2,   \qquad 
S^{D3}_{0}=\int_{{\cal M}^{1+3}}  {\cal L}^{D3}_4,   
\end{equation} 
over the the worldsheet ${\cal M}^{1+1}=  
\left( \xi^{\mu} \right)= \left( \tau , \s \right)$  and the 
D3-brane worldvolume ${\cal M}^{1+3}= {\zeta^m}$ ($m=0,\ldots , 3$).
 They should be regarded as surfaces embedded into the 
$D=10$ type $IIB$ superspace $\underline{{\cal M}}^{(1+9|32)}$
\begin{equation}\plabel{IIBss}
\underline{{\cal M}}^{(1+9|32)}= \{ X^{\underline{m}},
\Theta^{\underline{\mu}1}, \Theta^{\underline{\mu}2} \}
=\{ X^{\underline{m}},
\Theta^{\underline{\mu}I} \} , 
\qquad {\underline{m}}=0,\ldots ,9 , 
\quad {\underline{\mu}}=1,\ldots ,16, \quad I=1,2,   
\end{equation} 
\begin{equation}\plabel{IIBst}
{\cal M}^{1+1}\rightarrow \underline{{\cal M}}^{(1+9|32)}: \quad 
X^{\underline{m}} 
= \hat{X}^{\underline{m}} (\xi^{\mu} ), \quad 
\Theta^{I \underline{\mu}} = \hat{\Theta}^{I \underline{\mu}} 
(\xi^{\mu }),
\quad 
\end{equation}
\begin{equation}\plabel{IIBD3}
{\cal M}^{1+3}\rightarrow \underline{{\cal M}}^{(1+9|32)}: \quad 
X^{\underline{m}} 
= \tilde{X}^{\underline{m}} (\zeta^m ), \quad 
\Theta^{I \underline{\mu}} = \tilde{\Theta}^{I \underline{\mu}} 
(\zeta^m). 
\quad 
\end{equation}

The action for the interacting system 
is expected to be 
\begin{equation}\plabel{Sint0}
S= \int_{{\cal M}^{1+1}}  {\cal L}^{IIB}_2 +
\int_{{\cal M}^{1+3}}  {\cal L}^{D3}_4  + \int_{\partial {\cal M}^{1+1}} A ,  
\end{equation}
where the last term \cite{Mourad,Sezgin25} 
describes the interaction of the string endpoints 
with  the gauge field $A=dx^m A_m (x)$ of the super-D3-brane 
\cite{c1}. 
However, 
the action \p{Sint0}
immediately poses a problem. Its origin is 
the sum of 
integrals over different manifolds, which could 
have some nontrivial intersections, e.g.  
 \begin{equation}\plabel{inters}
\partial {\cal M}^{1+1} \in {\cal M}^{1+3}  \qquad \rightarrow \qquad  
{\cal M}^{1+3} \cap {\cal M}^{1+1} = \partial {\cal M}^{1+1} .
 \end{equation}

The intersection manifold $\partial {\cal M}^{1+1}$
for the superstring(s)---super--D3--brane system is a set of worldlines 
which may be numbered by a label $j$ 
(connected components of the superstring worldsheet boundary). 
For the case of one 
open superstring ending on the D3--brane $j =1,2$.  
Each  worldline can be parametrized by the 
proper time $\tau_j$.  
Thus the manifold $\partial {\cal M}^{1+1}$ 
can be defined parametrically  as a submanifold of  
 the worldsheet ${\cal M}^{1+1}$ 
 \begin{equation}\plabel{M1M11}
\partial {\cal M}^{1+1} ~~\in ~~{\cal M}^{1+1}: \qquad 
\xi^{\mu} = \xi^{\mu} (\tau _j ),
\end{equation}
and as a submanifold of the worldvolume 
${\cal M}^{1+3}$
 \begin{equation}\plabel{M1M13}
\partial {\cal M}^{1+1} ~~\in ~~{\cal M}^{1+3}: \qquad 
\zeta^{m} = \zeta^{m} (\tau _j ).
\end{equation}
In the following, for simplicity, we skip the index $j $. 
The bosonic and fermionic coordinate functions \p{IIBst}, 
\p{IIBD3} which define the embeddings of the worldsheet and the worldvolume 
into the target superspace should 
have the same image on   $\partial {\cal M}^{1+1}$ 
(i.e. should 
coincide when restricted to 
 $\partial {\cal M}^{1+1}$) 
\begin{equation}\plabel{bc}
\hat{X}^{\underline{m}} \left(\xi^{\mu} (\tau) \right)
= \tilde{X}^{\underline{m}}  \left(\zeta^m  (\tau) \right) 
\equiv \hat{\tilde{X}}^{\underline{m}} (\tau), 
\qquad 
\hat{\Theta}^{I \underline{\mu}} \left(\xi^{\mu} (\tau) \right)= 
\tilde{\Theta}^{I \underline{\mu}}  \left(\zeta^m  (\tau) \right)
\equiv \hat{\tilde{\Theta}}^{I \underline{\mu}} (\tau).
\qquad 
\end{equation} 

The above mentioned problem with the action \p{Sint0}
 appears because, due to the identification 
\p{bc}, the variations 
$\d \hat{X}^{\underline{m}} \left(\xi^{\mu} \right), \d 
\hat{\Theta}^{I \underline{\mu}} \left(\xi^{\mu} \right)$ 
and  $\d \tilde{X}^{\underline{m}}  \left(\zeta^m  \right), 
\d \tilde{\Theta}^{I \underline{\mu}}  \left(\zeta^m  \right)$
may not be regarded as completely independent ones. 

The straightforward way to take Eqs. \p{bc} into account 
is to incorporate them into the action \p{Sint0} by means of 
a bosonic and a fermionic Lagrangian multiplier 1-form 
$\hat{\tilde{P}}_{1\underline{m}} = d\tau 
\hat{\tilde{P}}_{\tau \underline{m}}$ 
and $\hat{\tilde{\pi}}_{1\underline{\mu}} = 
d\tau \hat{\tilde{\pi}}_{\tau \underline{\mu}}$ 
\begin{equation}\plabel{Sint01}
S^s = \int_{{\cal M}^{1+1}}  {\cal L}^{IIB}_2 +
\int_{{\cal M}^{1+3}}  {\cal L}^{D3}_4  + \int_{\partial {\cal M}^{1+1}} A + 
\end{equation} 
$$
+ \int_{\partial {\cal M}^{1+1}} \hat{\tilde{P}}_{1\underline{m}}  
\left( \hat{X}^{\underline{m}} \left(\xi^{\mu} (\tau) \right)
-  \tilde{X}^{\underline{m}}  \left(\zeta^m  (\tau) \right) \right)
+ \int_{\partial {\cal M}^{1+1}} i \hat{\tilde{\pi}}_{1\underline{\mu}} 
\left( 
\hat{\Theta}^{I \underline{\mu}} \left(\xi^{\mu} (\tau) \right)-  
\tilde{\Theta}^{I \underline{\mu}}  \left(\zeta^m  (\tau) \right)
\right).  
$$  
However, 
as we will see below,  these Lagrange multipliers 
cannot be determined from the equations of motion. 
Because, in addition,  
 their nature may seem unclear,  
  doubts could  arise  
whether the Lagrange multiplier method is applicable at all here.  
An example of a system where 
this method indeed fails is provided by 
 self-dual gauge fields, whose covariant description 
at the level of the action functional 
required the development of a  
 special (PST) approach \cite{PST}.

Thus 
 to justify the applicability of the action 
\p{Sint01} it is useful 
to describe 
the interacting branes in a different manner.

In fact, 
in order to be able to vary the action \p{Sint0} directly, one 
 could try (instead of using Lagrange multipliers \p{Sint01}) 
to find a 
{\sl supersymmetric} way to 
write 
all the terms as integrals over a manifold containing both 
the super-D3-brane worldvolume  and the superstring worldsheet. 
 
It turns out that this is indeed possible.
Moreover, the dynamical system then may be extended possibly 
by inclusion of an action for supergravity.

\section{Space-time filling branes and induced embeddings}

We find it useful to first extend 
our system by inclusion 
of the super-D9-brane, which is a {\sl space-time filling brane} of the 
$D=10$ type $IIB$ superspace  
\begin{equation}\plabel{Sint1}
S= \int_{{\cal M}^{1+9}}  {\cal L}_{10} +
\int_{{\cal M}^{1+1}}  {\cal L}^{IIB}_2 +
\int_{{\cal M}^{1+3}}  {\cal L}^{D3}_4  + \int_{\partial {\cal M}^{1+1}} A .  
\end{equation}
Here ${\cal L}_{10}$ is the super-D9-brane Lagrangian form 
(see e.g. \cite{abkz}). 
The essential point is that the super-D9-brane 
implies the existence 
of the map of a $d=10$ dimensional bosonic surface 
${\cal M}^{1+9}= \{ x^{\bar{m}} \}$ ($\bar{m}=0,\ldots, 9 $) into  
type $IIB$ superspace 
\begin{equation}\plabel{IIBD9}
{\cal M}^{1+9}\rightarrow \underline{{\cal M}}^{(1+9|32)}: \quad 
X^{\underline{m}} 
= \bar{X}^{\underline{m}} (x^{\bar{m}}), \quad 
\Theta^{ I \underline{\mu}} = \bar{\Theta}^{I \underline{\mu}} 
(x^{\bar{m}}),
\end{equation}
with an {\sl invertible function} 
$X^{\underline{m}} 
= \bar{X}^{\underline{m}} (x^{\bar{m}})$. 
This allows 
the definition of an induced embedding of the superstring worldsheet 
and the super-D3-brane worldvolume into the bosonic surface 
${\cal M}^{1+9}$
(D9-brane worldvolume) 
\begin{equation}\plabel{indIIB}
x^{\bar{m}}=\hat{x}^{\bar{m}}(\xi ) \quad \leftarrow \quad  
\hat{X}^{\underline{m}} (\xi )
= \bar{X}^{\underline{m}} \left(\hat{x}^{\bar{m}}(\xi) \right),
\end{equation}
\begin{equation}\plabel{indD3} 
x^{\bar{m}}=\tilde{x}^{\bar{m}}(\zeta ) \quad  \leftarrow \quad  
\tilde{X}^{\underline{m}} (\zeta ) 
= \bar{X}^{\underline{m}} \left(\tilde{x}^{\bar{m}}(\zeta) \right), 
\end{equation}
and to consider the superstring and super-D3-brane coordinate functions 
as images of the functions defined on ${\cal M}^{1+9}$ on the 
 worldsheet and on the worldvolume, respectively: 
\begin{equation}\plabel{IIBstD9}
\hat{X}^{\underline{m}} (\xi ) 
= \bar{X}^{\underline{m}} \left(\hat{x}^{\bar{m}}(\xi) \right), \quad 
\hat{\Theta}^{I \underline{\mu}} 
(\xi )= 
\bar{\Theta}^{I \underline{\mu}} \left(\hat{x}^{\bar{m}}(\xi) \right),
\quad 
\end{equation}
\begin{equation}\plabel{IIBD3D9}
\tilde{X}^{\underline{m}} (\zeta ) 
= \bar{X}^{\underline{m}} \left(\tilde{x}^{\bar{m}}(\zeta)\right), 
 \qquad 
\tilde{\Theta}^{I\underline{\mu}} 
(\zeta^m)
= \bar{\Theta}^{I \underline{\mu}}  \left(\tilde{x}^{\bar{m}}(\zeta)
\right).
\qquad 
\end{equation}

Actually, what we need are the induced embeddings 
\p{IIBstD9}, \p{IIBD3D9}. Below we will treat the 9-brane as auxiliary 
and drop the Lagrangian ${\cal L}_{10}$ altogether.
The study of the interaction of the  fundamental string with 
this super-D9-brane 
in the framework of the present approach is the subject of 
another paper \cite{BK}. 

An interesting alternative for future study would be 
to consider ${\cal L}_{10}$ in \p{Sint1} 
as a Lagrangian form of  a counterpart of 
the group manifold action 
\cite{grm} for 
$D=10$ type $IIB$ supergravity, 
which  assumes 
the map \p{IIBD9} as well. 
This provides the possibility to 
generalize our consideration for the case of curved 
superspace. 
Thus the  construction of such group manifold action for type $IIB$ 
supergravity
on the basis of 
the PST action \cite{lst} seems to be another promising problem, which we 
however do not address here.


\section{Current form distributions and supersymmetry}

\subsection{Covariant current distribution forms}

To write the action for the coupled system \p{Sint1} in a unique form, let us 
define first the 10-dimensional manifestly covariant 
current densities with  support on the superstring worldsheet 
and on the super-D3-brane worldvolume respectively 
(see \cite{bbs} for the bosonic M2-brane and M5-brane \cite{blnpst}
interacting with D=11 supergravity) 
\begin{equation}\plabel{J81}
J_8 = (dx)^{\wedge 8}_{\bar{n} \bar{m}} J^{\bar{n} \bar{m}} 
= (dx)^{\wedge 8}_{\bar{n} \bar{m}} 
\int_{{\cal M}^{1+1}} 
d\hat{x}^{\bar{m}} (\xi ) \wedge d\hat{x}^{\bar{n}} (\xi )
\d^{10} \left( x - \hat{x} (\xi )\right) , 
\end{equation}
\begin{equation}\plabel{J61}
J_6 = (dx)^{\wedge 6}_{\bar{m}_1\ldots  \bar{m}_4} 
J^{\bar{m}_1\ldots  \bar{m}_4} =
(dx)^{\wedge 6}_{\bar{m}_1\ldots  \bar{m}_4} 
\int_{{\cal M}^{1+3}} 
d\tilde{x}^{\bar{m}_1} (\zeta ) \wedge \ldots \wedge 
d\tilde{x}^{\bar{m}_4} (\zeta )
\d^{10} \left( x - \tilde{x} (\zeta )\right) , 
\end{equation}
with 
\begin{equation}\plabel{dxn}
(dx)^{\wedge n}_{\bar{m}_1\ldots  \bar{m}_{10-n}} \equiv 
{ 1 \over n! (10-n)!} \e_{\bar{m}_1\ldots  \bar{m}_{10-n} 
\bar{n}_1 \ldots \bar{n}_n}  
d{x}^{\bar{n}_1 } \wedge \ldots \wedge
dx^{\bar{n}_n} 
\end{equation}
Their main properties are  
\begin{equation}\plabel{J8def1}
\int_{{{\cal M}}^{1+9}} J_8 \wedge {\cal L}_2 = 
\int_{{\cal M}^{1+1}} \hat{{\cal L}}_2 , ~~~ \qquad ~~~ 
\int_{{{\cal M}}^{1+9}} J_6 \wedge {\cal L}_4 = 
\int_{{\cal M}^{1+1}} \tilde{{\cal L}}_4 ,  
\end{equation}
where 
\begin{equation}\plabel{L2D9}
 {\cal L}_2 = {1 \over 2} dx^{\bar{m}} \wedge dx^{\bar{n}}  
{\cal L}_{\bar{n} \bar{m}} ( x ), 
\qquad 
 {\cal L}_4 = {1 \over 4!} dx^{\bar{m}_1} \wedge \ldots \wedge dx^{\bar{m}_4}  
{\cal L}_{\bar{m}_4\ldots  \bar{m}_1} ( x ) 
\qquad 
\end{equation}
are arbitrary 2-form and 4-forms {\sl on the 
bosonic surface} ${{\cal M}}^{1+9}$ and 
\begin{equation}\plabel{L2D9s}
\hat{{\cal L}}_2 = {1 \over 2}
 d\hat{x}^{\bar{m}} (\xi ) \wedge d\hat{x}^{\bar{n}} (\xi )
{\cal L}_{\bar{n}\bar{m}} ( \hat{x}^{\bar{l}}(\xi)),  
\qquad 
\tilde{{\cal L}}_4 = {1 \over 4! }
 d\tilde{x}^{\bar{m}_1} (\zeta ) \wedge\ldots \wedge 
d\tilde{x}^{\bar{m}_4} (\zeta )
{\cal L}_{\bar{m}_4\ldots  \bar{m}_1} ( \tilde{x}^{\bar{l}}(\zeta)) 
\end{equation}
are their pull-backs onto the worldsheet and the worldvolume, respectively.

\subsection{Brane boundary and current (non)conservation}

If we assume that the worldvolume of a brane, say super-D3-brane, is closed
$\partial {\cal M}^{1+3}= 0$, then the brane current 
$J^{\bar{m}_1... \bar{m}_4}$ \p{J61} is conserved, 
$\partial_{\bar{m}_4}J^{\bar{m}_1... \bar{m}_4}=0$, 
and thus the  
 current density form $J_6$  \p{J61} is a closed form $dJ_6=0$. 

This is not true for open branes. 
Here we are interested in the open superstring case $\partial {\cal M}^{1+1}= 
\{ \tau \} \not= 0$. 
Substituting  
the closed two form, say $dA$, 
 instead of ${\cal L}_2$ 
into \p{J8def1} 
and using Stokes' theorem one arrives at 
\begin{equation}\plabel{dJ8def}
\int_{\partial {\cal M}^{1+1}} A = 
\int_{{\cal M}^{1+1}} dA = 
\int_{{{\cal M}}^{1+9}} J_8 \wedge dA = 
\int_{{{\cal M}}^{1+9}} dJ_8 \wedge A . 
\end{equation}
Eq. \p{dJ8def} demonstrates that 
the form $dJ_8$ has a support localized at the boundary of the worldsheet, 
i.e. on the worldline of the string endpoints parametrized by the proper time 
$\tau$. This again can be justified by an explicit calculation with Eqs. 
\p{J81}, which results in 
\begin{equation}\plabel{dJ80}
dJ_8 = 
- (dx)^{\wedge 9}_{\bar{n}} 
\int_{\partial{\cal M}^{1+1}} 
d\hat{x}^{\bar{n}} (\tau ) 
\d^{10} \left( x - \hat{x} (\tau )\right) . 
\end{equation}

For the description of the above  situation 
it is useful to introduce also 
superstring and super-D3-brane 
current form distributions $j_1$ and $j_3$ 
with support on the boundary of the superstring worldsheet 
\p{inters} 
\begin{equation}\plabel{j10}
j_1 = d\xi^{\mu} \e_{\mu\nu} \int_{\partial {\cal M}^{1+1}} 
d\tilde{\xi}^{\nu} (\tau ) 
\d^{2} \left( \xi - \tilde{\xi} (\tau )\right)~, \qquad 
\xi^{\mu}=(\tau , \s ) 
\end{equation} 
\begin{equation}\plabel{j30}
j_3 = d\zeta^{\wedge 3}_m \int_{\partial {\cal M}^{1+1}} 
d\hat{\zeta}^{m} (\tau ) 
\d^{4} \left( \zeta - \hat{\zeta} (\tau )\right)~, \qquad 
\zeta^m = (\zeta^0,\ldots , \zeta^3) 
\end{equation} 
with the properties 
\begin{equation}\plabel{j1def}
\int_{{\cal M}^{1+1}} j_1 \wedge 
\hat{A} = \int_{\partial {\cal M}^{1+1}} \hat{A},   
~~~  \qquad ~~~
 \int_{{\cal M}^{1+3}} j_3 \wedge 
\tilde{A} = \int_{\partial {\cal M}^{1+1}} \hat{A}.  
\end{equation}
Collecting Eqs. \p{j1def}  and \p{dJ80} 
\begin{equation}\plabel{dJ8i}
\int_{{\cal M}^{1+1}} d\hat{A} = 
\int_{{\cal M}^{1+9}} dJ_8 \wedge {A} 
= \int_{{\cal M}^{1+3}} j_3 \wedge 
\tilde{A} = \int_{{\cal M}^{1+1}} j_1 \wedge 
\hat{A} 
\end{equation}
one can write down formal  
 relations between current distribution forms  
\begin{equation}\plabel{dJ8f}
dJ_8 = j_1 \wedge J_8  = j_3 \wedge J_6 
\end{equation} 
Here  formal extrapolations of the relations 
\p{J8def1} to arbitrary worldvolume forms 
have been used. 

Eq. \p{dJ8f} represents in a compact and transparent manner 
 the nonconservation 
of the superstring current form due to the presence of the worldsheet 
boundary. Let us stress, however, that it is really true in the sense of 
the integrated equations \p{dJ8i} with a test 1-form $A$.

\subsection{Supersymmetric invariance of distribution forms}

Performing a general coordinate transformation, 
the current densities can be expressed  in terms of the 
coordinate fields $X^{\underline{m}}$ as, e.g. 
 \begin{equation}\plabel{J80}
J_8 = (dX)^{\wedge 8}_{\underline{n} \underline{m}} 
J^{\underline{n} \underline{m}}( X) = 
(dX)^{\wedge 8}_{\underline{n} \underline{m}} 
\int_{{\cal M}^{1+1}} 
d\hat{X}^{\underline{m}} (\xi ) \wedge d\hat{X}^{\underline{n}} (\xi )
\d^{10} \left( X - \hat{X} (\xi )\right)~.
\end{equation} 
In \p{J80} one recognizes the current densities used for the description 
of intersection of the bosonic branes \cite{bbs}. It does not include 
the Grassmann coordinate fields $\Theta^{I}, \hat{\Theta}^{I}$ 
and this may lead to a doubts  concerning  its invariance 
under supersymmetry, which, however, holds 
after the identification \p{IIBstD9},  \p{IIBD3D9}, as will be seen below. 

The variation of the form 
\p{J80} 
can be written as 
\begin{equation}\plabel{deJ80}
\d J_8 = 
3 (dX)^{\wedge 8}_{[\underline{m} \underline{n}} \partial_{\underline{k}]}
\int_{{\cal M}^{1+1}} 
d\hat{X}^{\underline{m}} (\xi ) \wedge d\hat{X}^{\underline{n}} (\xi )
\left( \d X^{\underline{k}}- \d \hat{X}^{\underline{k}} (\xi ) \right)~ 
\d^{10} \left( X - \hat{X} (\xi )\right)~- 
\end{equation}
$$
- 2 (dX)^{\wedge 8}_{\underline{m} \underline{n}}
\int_{\partial {\cal M}^{1+1}} 
d\hat{X}^{\underline{m}} (\tau ) \left( \d X^{\underline{n}}- 
\d \hat{X}^{\underline{n}} (\tau ) \right)~
\d^{10} \left( X - \hat{X} (\tau )\right)~.
$$

The target superspace supersymmetry transformations 
\begin{equation}\plabel{susy1}
\d X^{\underline{m}} = \Theta^{I} \sigma^{\underline{m}} \e^I , \qquad  
\d \Theta^{I\underline{\mu}} =  \e^{I\underline{\mu}} 
\end{equation}
imply the transformations of the superstring coordinate fields 
\begin{equation}\plabel{susyIIB}
\d \hat{X}^{\underline{m}}(\xi ) = \hat{\Theta}^{I} (\xi ) 
\sigma^{\underline{m}} \e^I , 
\qquad  
\d \hat{\Theta}^{I\underline{\mu}}(\xi )  =  \e^{I\underline{\mu}},  
\end{equation}
and the ones of the 
(auxiliary) 9-brane   
\begin{equation}\plabel{susyD9}
\d X^{\underline{m}} (x) = \Theta^{I} (x) \sigma^{\underline{m}} \e^I , 
\qquad  
\d \Theta^{I\underline{\mu}} (x) =  \e^{I\underline{\mu}}.  
\end{equation}

In the parametrization of the 9-brane worldvolume by  $X^{\underline{m}}$
coordinates, which is possible due to the invertibility \p{IIBD9} of the 
embedding function 
$X(x)$, 
the transformation 
\p{susyD9} coincides with the Goldstone fermion realization \cite{VA} 
of the type $IIB$ supersymmetry 
\begin{equation}\plabel{susyD9X}
\d X^{\underline{m}} = \Theta^{I} (X) \sigma^{\underline{m}} \e^I , \qquad  
\d \Theta^{I\underline{\mu}} (X) \equiv \Theta^{I\underline{\mu}~\prime} 
(X^{\prime} )- \Theta^{I\underline{\mu}} (X)
= \e^{I\underline{\mu}} . 
\end{equation}

The variation of the current form \p{J80} under the transformations 
 \p{susyIIB} (cf. \p{deJ80}) becomes  
\begin{equation}\plabel{deJ80susy}
\d J_8 = 
3 (dX)^{\wedge 8}_{[\underline{m} \underline{n}} \partial_{\underline{k}]}
\int_{{\cal M}^{1+1}} 
d\hat{X}^{\underline{m}} (\xi ) \wedge d\hat{X}^{\underline{n}} (\xi )
\left( \Theta^{I} (X) 
- \hat{\Theta}^{I} (\xi ) \right)\sigma^{\underline{k}} \e^I ~
\d^{10} \left( X - \hat{X} (\xi )\right)~- 
\end{equation}
$$
- 2 (dX)^{\wedge 8}_{\underline{m} \underline{n}}
\int_{\partial {\cal M}^{1+1}} 
d\hat{X}^{\underline{m}} (\tau ) \left( \Theta^{I} (X) 
- \hat{\Theta}^{I} 
(\tau ) \right)\sigma^{\underline{n}} \e^I ~
\d^{10} \left( X - \hat{X} (\tau )\right)~.
$$

The key observation is that if one identifies the 
 Grassmann coordinates fields of the lower dimensional branes 
$\hat{\Theta}^{I} (\xi)$,  $\tilde{\Theta}^{I} (\zeta)$   
with the images of the 
9-brane Grassmann coordinate fields on the worldvolumes 
(Goldstone fermions \cite{VA}) ${\Theta}^{I} (X)$ 
\begin{equation}\plabel{Thid1}
\hat{\Theta}^{I} (\xi)= {\Theta}^{I} \left(\hat{X}(\xi)\right), 
\qquad 
\tilde{\Theta}^{I} (\zeta)= {\Theta}^{I} \left(\tilde{X}(\zeta )\right)
\end{equation}
one finds that 
{\sl the current forms $J_8$ and $J_6$ are supersymmetric invariant}! 

Such an invariance is quite evident in the representation 
\p{J81}, \p{J61}, as the 
coordinates $x^n$ \p{IIBD9} are 
inert under the target space supersymmetry. 
The identification \p{Thid1} 
(see \p{indIIB}, \p{indD3})
\begin{equation}\plabel{Thid}
\hat{\Theta}^{I} (\xi)= {\Theta}^{I} \left(\hat{x}(\xi)\right), 
\qquad 
\tilde{\Theta}^{I} (\zeta)= {\Theta}^{I} \left(\tilde{x}(\zeta )\right)
\end{equation}
is implied here by the assumption that it is 
 possibile to lift the complete superbrane actions 
to the 10-dimensional integral form  using the relations 
\p{J8def1}.  
The manifestly supersymmetric form of the current densities 
appears after passing to the 
supersymmetric   basis of the  space tangent to 
${\cal M}^{1+9}$ 
\begin{equation}\plabel{Pipb}
\Pi^{\underline{m}}= 
 dx^m  \Pi^{~\underline{m}}_{{m}} 
= d {X}^{\underline{m}}
- i 
d{\Th}^I \s^{\underline{m}} {\Th}^I, \qquad  
\Pi^{~\underline{m}}_{{n}}
\equiv 
\partial_n {X}^{\underline{m}}
- i \partial_{{n}} 
{\Th}^I \s^{\underline{m}} {\Th}^I.  
\end{equation}
We arrive at 
\begin{equation}\plabel{J8Pi8}
J_8 = 
(\Pi)^{\wedge 8}_{\underline{n} \underline{m}} 
{ 1 \over det(\Pi_{{r}}^{~\underline{s}})} 
\int_{{\cal M}^{1+1}} 
\hat{\Pi}^{\underline{m}} \wedge \hat{\Pi}^{\underline{n}} 
\d^{10} \left( x - \hat{x} (\xi )\right), 
\end{equation}
\begin{equation}\plabel{J6Pi6}
J_6 = 
(\Pi)^{\wedge 6}_{\underline{m}_1 \ldots \underline{m}_4} 
{ 1 \over det(\Pi_{{r}}^{~\underline{s}})} 
\int_{{\cal M}^{1+1}} 
\tilde{\Pi}^{\underline{m}_1} \wedge \ldots 
\wedge \tilde{\Pi}^{\underline{m}_4} 
\d^{10} \left( x - \tilde{x} (\zeta )\right).
\end{equation}
 
\section{Lagrangian forms and action for the interacting system}

Thus, {\sl if we assume that the Lagrangian form  for superstring 
${\cal L}_{2}^{IIB}$
and super-D3-brane ${\cal L}_{4}^{D3}$
can be presented as the pull-back of some 10-dimensional 
2-form and 4-form living on the bosonic surface} ${\cal M}^{1+9}$ 
(see \p{L2D9}, \p{L2D9s}), we can write 
the action for the coupled system \p{Sint0} in a  way which allows  
 direct variation 
\begin{equation}\plabel{Sint}
S^G= \int_{{\cal M}^{1+9}}  \left[
J_8 \wedge {\cal L}^{IIB}_2 + 
J_6 \wedge {\cal L}^{D3}_4 + dJ_8 \wedge A  
\right]
\end{equation}

The above requirement is {\sl not} satisfied 
by the leading (kinetic) terms of  the standard actions 
\cite{gsw,c1} 
\begin{equation}\plabel{SstIIB}
{\cal L}^{IIB}_2 = d^2 \xi 
\sqrt{det(\hat{g}_{\mu\nu})
)} 
- B_2, \quad 
\end{equation}
\begin{equation}\plabel{SstD3}
{\cal L}^{D3}_4 = d^4 \zeta  
\sqrt{det(g_{mn} + {\cal F}_{mn})} + e^{\cal{F}} \wedge  C~ \vert_{4} 
\end{equation}
where 
\begin{equation}\plabel{gind}
\hat{g}_{\mu\nu}\equiv \hat{\Pi}^{\underline{m}}_\mu 
\hat{\Pi}_{\underline{m}\nu}, 
\quad \hat{\Pi}^{\underline{m}}= d\xi^\mu \hat{\Pi}^{\underline{m}}_\mu, 
\qquad 
g_{mn} = \tilde{\Pi}^{\underline{m}}_m \tilde{\Pi}_{\underline{m}n}, 
\quad \tilde{\Pi}^{\underline{m}}= d\zeta^m \tilde{\Pi}^{\underline{m}}_m
\end{equation}
are the superstring and the super--D3--brane induced metrics 
and ${\cal F}$ is the generalized field strength of the gauge field $A$
\begin{equation}\plabel{calF}
{\cal F} = dA -B_2
\end{equation}
The D3--brane Wess-Zumino term is defined by \cite{c1}
\begin{equation}\plabel{WZD3}
e^{\cal{F}} \wedge  C~ \vert_{4}
= C_4 + C_2 \wedge {\cal F} + C_0 \wedge {\cal F} \wedge {\cal F}. 
\end{equation} 
Here $C_{2k}$ are RR gauge fields of type $IIB$ supergravity 
with a flat superspace field strength 
\begin{equation}\plabel{RRR}
R = \oplus _{n=0}^{5} R_{2n+1} =  
 e^{- {\cal F}} \wedge   d( e^{\cal{F}} \wedge  C) 
= 2i d\Theta^{2\underline{\nu} }  \wedge  d\Theta^{1\underline{\mu} } \wedge 
\oplus _{n=0}^{4} \hat{\s}^{(2n+1)}_{\underline{\nu}\underline{\mu} }, \qquad 
\end{equation}
while 
$B_2$, entering \p{SstIIB}, \p{calF},  is the NS-NS gauge field, whose 
flat superspace value 
is 
\begin{equation}\plabel{B2def}
B_2 = i \Pi^{\underline{m}}\wedge 
\left(
d\Th^1\s_{\underline{m}}\Th^1  - d\Th^2\s_{\underline{m}}\Th^2 
\right) + 
d\Th^1\s^{\underline{m}}\Th^1 \wedge  d\Th^2\s_{\underline{m}}\Th^2 
\end{equation} 
\begin{equation}\plabel{H3def}
H_3= dB_2 = i \Pi^{\underline{m}}\wedge 
\left(
d\Th^1\s_{\underline{m}} \wedge d\Th^1  - d\Th^2\s_{\underline{m}}\wedge 
d\Th^2 \right). 
\end{equation}

One can actually consider the action \p{Sint} with Lagrangian 
form \p{SstIIB}, \p{SstD3}, extending formally 
the relations \p{J8def1} to arbitrary forms on the 
worldsheet and worldvolume respectively. 
However, a more rigorous procedure 
(which actually  
could motivate formal manipulations of this type 
also in another context) consists in searching for 
an equivalent representation of the superstring and superbrane 
actions, whose Lagrangian form can be considered as a pull--back of the 
10-dimensional forms. 
Fortunately such actions do exist.
They were proposed in the frame 
of the Lorentz harmonic approach 
for superstrings \cite{BZ} 
(see also \cite{bpstv,bsv,baku}) and 
super-Dp-branes  
\cite{bst,baku,abkz} respectively.

Thus in the  Lorentz harmonic approach the action \p{Sint}   
of the interacting system of the super-D3-brane and 
the fundamental superstring 
ending on the super--D3--brane becomes 
\begin{eqnarray}\plabel{SD3+IIB} 
&& S^G = \int\limits_{{\cal{M}}^{1+9}}^{}  
J_6 \wedge \Big[E^{\wedge 4} \sqrt{-det(\eta_{ab} 
+ F_{ab})} + 
Q_{2} \wedge \left( dA-B_2  - {1 \over 2} 
E^a \wedge 
E^b ~F_{ba} \right)  
+ e^{\cal{F}} \wedge  C~ \vert_{4} ~\Big] + 
\nonumber \\ 
&& + \int_{{\cal{M}}^{10}} J_8 \wedge \left({1 \over 2} 
{E}^{++} \wedge {E}^{--}   
- B_{2} \right)  +  \int_{{\cal{M}}^{1+9}} dJ_8 \wedge A . 
\end{eqnarray} 
Here $Q_2$ is a 2-form Lagrange multiplier, $F_{ab}=-F_{ba}$ 
is an auxiliary $d=4$ antisymmetric tensor field  and 
\begin{equation}\plabel{Ea}
{E}^{a} = {\Pi}^{\underline{m}} 
{u}_{\underline{m}}^{a},  
\qquad     
{E}^{\pm\pm} = {\Pi}^{\underline{m}} 
U_{\underline{m}}^{\pm\pm},  
\qquad     
\end{equation}
where   
\begin{equation}\plabel{harmvD3}
{u}^{\underline{a}}_{\underline{m}} (\xi ) 
\equiv ( 
{u}^{a}_{\underline{m}}, 
{u}^{i}_{\underline{m}}
) \quad \in \quad SO(1,D-1)\qquad 
\end{equation} 
$$ 
\Rightarrow  
 {u}^{a}_{\underline{m}}{u}^{b\underline{m}}= \eta^{ab}, ~~ 
{u}^{a}_{\underline{m}} 
{u}^{i\underline{m}} =0, \ldots , \qquad   
 i=1,\ldots, 6 $$
\begin{equation}\plabel{harmvIIB}
U^{\underline{a}}_{\underline{m}} (\xi ) 
\equiv ( U^{++}_{\underline{m}}, U^{--}_{\underline{m}}, 
U^{\hat{i}}_{\underline{m}}) \quad \in \quad SO(1,D-1) \qquad 
\end{equation} 
$$ 
\Rightarrow 
  U^{++}_{\underline{m}} U^{++\underline{m}}=0, \qquad   
U^{--}_{\underline{m}} U^{--\underline{m}}=0, \qquad 
U^{\hat{i}}_{\underline{m}} U^{\pm\pm\underline{m}}=0,
~\ldots , \hat{i}=1,\ldots , 8 
$$
are auxiliary Lorentz group valued matrix fields 
(Lorentz harmonics, see \cite{sok,BZ,bpstv,bsv} and refs. in \cite{bpstv}).

 \section{Properties of the equations of motion for coupled branes} 

Our study so far relies on the embedding of the branes into the 
Goldstone fermion theory (dynamical or auxiliary  (D-1)--brane)
rather than into the superspace. Thus we clearly  
have less fermionic equations for the coupled branes 
than can be expected. 
Nevertheless, as we will see, this is not a drawback of 
just the GFE action \p{Sint}, \p{SD3+IIB}, 
as the equations obtained 
from the 
SSPE action 
 \p{Sint01} are actually equivalent.

For the 
SSPE action 
\p{Sint01} we have twice as many  
fermionic variables as in the case above. However it includes  
 the Lagrange multiplier 1-forms 
$P_1$, $\pi_1$ which remain indefinite and 
appear in the equations of motion just inside a source 
localized at the intersection. 
As we clarify below, these equations with indefinite source 
are equivalent to the fermionic equations following from the 
GFE action.

We first consider 
the SSPE action \p{Sint01} with 
\begin{equation}\plabel{LIIB}
{\cal L}^{IIB}_2 = {1 \over 2} 
{E}^{++} \wedge {E}^{--}   
- B_{2}
\qquad     
\end{equation}
\begin{equation}\plabel{LD3}
 {\cal L}^{D3}_4 = E^{\wedge 4} \sqrt{-det(\eta_{ab} 
+ F_{ab})} + 
Q_{2} \wedge \left( dA-B_2  - {1 \over 2} 
E^a \wedge 
E^b ~F_{ba} \right)  
+ e^{\cal{F}} \wedge  C~ \vert_{4} 
\qquad     
\end{equation}
\cite{bsv,bst} (see \p{SD3+IIB}--\p{harmvIIB}). 
To obtain the simplest form of the equations of motion 
it is convenient to pass from the  'holonomic' basis in the space of 
variations
$\d X, \d \Theta^I , \d A, \ldots $ to the 
'supersymmetric' one (cf. \p{Pipb}, \p{calF}, \p{B2def}, \p{Ea}, \p{harmvD3}, 
\p{harmvIIB}) 
\begin{equation}\plabel{varba}
i_\d \Pi^{\underline{m}} 
= \d {X}^{\underline{m}}
- i \d{\Th}^I \s^{\underline{m}} {\Th}^I, \qquad   
\d{\Th}^{I\underline{\mu}}, \qquad 
i_\d ({\cal F}-F) \equiv \d A - i_\d B_2 + E^b F_{ba} i_\d E^a , \quad \ldots .
\end{equation}
Then the variation with respect to coordinate fields 
$\hat{X}(\xi), \hat{\Theta}^{I}(\xi)$ 
and $\tilde{X}(\zeta), \tilde{\Theta}^{I}(\zeta)$ becomes 
 \begin{equation}\plabel{deSs}
\d S^{s} = 
\int_{{\cal M}^{1+1}}  
\left( 
(\hat{M}_{2\underline{m}} - j_1 \wedge P_{1\underline{m}}
+ j_1 \wedge \hat{\tilde{M}}_{1\underline{m}} ) ~
i_\d \hat{\Pi}^{\underline{m}} 
+ i 
(\hat{\Psi}^I_{2\underline{\mu}} 
- j_1 \wedge \psi_{1\underline{\mu}}) \d  
\hat{\Theta}^{I\underline{\mu}} 
\right) + 
\end{equation}
$$
+ \int_{{\cal M}^{1+3}}  
\left( (\tilde{M}_{4\underline{m}}  + j_3 \wedge P_{1\underline{m}}) 
i_\d \tilde{\Pi}^{\underline{m}} 
+ i ( 
\tilde{\Psi}^I_{2\underline{\mu}} +  j_1 \wedge \psi_{1\underline{\mu}})
\d \tilde{\Theta}^{I\underline{\mu}} 
\right),   
$$
where we used the density 1--forms $j_1, j_3$  \p{j1def} 
to lift the boundary inputs
to the worldsheet and the worldvolume, respectively,  
and abbreviate 
\begin{equation}\plabel{psi1}
\psi_{1\underline{\mu}}= \pi_{1\underline{\mu}} - i P_{1\underline{m}} 
(\s^{\underline{m}}  \hat{\tilde{\Theta}}^I)_{\underline{\mu}}.
\end{equation}
The expressions 
$\hat{M}_{2\underline{m}}$, $\tilde{M}_{4\underline{m}}$ and 
$\hat{\Psi}^I_{2\underline{\mu}}$, $\tilde{\Psi}^I_{4\underline{\mu}}$     
denote the l.h.s.-s. of the bosonic and fermionic equations 
for the free (closed) type $IIB$ superstring 
\begin{equation}\plabel{freeIIB}
\partial {\cal M}^{1+1}=0 ~~ \Rightarrow \qquad 
\hat{M}_{2\underline{m}}=0, \qquad 
\hat{\Psi}^I_{2\underline{\mu}}=0
\end{equation}
 and the free super--D3-brane  
\begin{equation}\plabel{freeD3}
\partial {\cal M}^{1+1}=0 ~~ \Rightarrow \qquad 
\tilde{M}_{4\underline{m}} =0, \qquad 
\tilde{\Psi}^I_{4\underline{\mu}}=0, 
\end{equation}
written in terms of differential forms \cite{bpstv,bst,baku,abkz} 
(we will not need their explicit expressions below), 
while 
$
\hat{\tilde{M}}_{1\underline{m}} 
$
denotes the coordinate variation localized at the boundary, which appears 
due to the integration by part in the 'bulk' superstring action 
\p{S0IIB}. 
We should stress that in the basis \p{varba} 
no boundary input 
with 
the variation $\d \Theta^I$  appears (see 
Appendix A, and \cite{BK} for details). 
Note also that we use 
the Lorentz harmonic formulations of superstring and super-D3-brane  
\cite{BZ,bpstv,baku} as here the free equations of motion appear  
in the form which allows a lifting to the 10-dimensional space,  
while the standard formulations \p{SstIIB}--\p{H3def} 
can be considered in such a way only formally. 

For the coupled system one can  expect 
some set of equations with sources localized at the intersection 
instead of \p{freeIIB}, \p{freeD3}.  

However, the fermionic equations which follows from \p{deSs} 
\begin{equation}\plabel{Theqs}
\hat{\Psi}_{2\underline{\mu}}^I  = j_1 \wedge \psi_{1\underline{\mu}}^I , 
\qquad  
\tilde{\Psi}_{4\underline{\mu}}^I  = - j_3 \wedge \psi_{1\underline{\mu}}^I , 
\qquad  
\end{equation}
include a source localized at the boundary and expressed through the 
the Lagrange multiplier 1-forms by \p{psi1}. This source
is {\sl indefinite}, as the Lagrange multipliers are not determined by the 
equations of motion.

In the above notations the variation of the 
GFE action 
\p{Sint}, \p{SD3+IIB} with respect to coordinate fields 
$X^{\underline{m}}(x)$ and $\Theta^{I\underline{\mu}}(x)$ 
reads as 
\begin{equation}\plabel{deSG}
\d S^{G}
= \int_{{\cal M}^{1+9}}  
\left( J_8 \wedge {M}_{2\underline{m}} 
+ J_6 \wedge M_{4\underline{m}} + dJ_8 \wedge M_{1\underline{m}}\right) 
i_\d \Pi^{\underline{m}} + 
\int_{{\cal M}^{1+9}}  \left( i J_8 \wedge {\Psi}^I_{2\underline{\mu}} 
+ i J_6 \wedge {\Psi}^I_{4\underline{\mu}} 
\right) \d {\Theta}^{I\underline{\mu}}  
\end{equation}
where the forms  $\hat{{M}}_{2\underline{m}}$,  
$\hat{\Psi}^I_{2\underline{\mu}}$,  
$ {\tilde{M}}_{4\underline{m}}$, $\tilde{\Psi}^I_{4\underline{\mu}}$
$\hat{\tilde{M}}_{1\underline{m}}$, 
 entering Eq. \p{deSs} 
are the pull--backs of the 10-dimensional forms 
${{M}}_{2\underline{m}}$, ${\Psi}^I_{2\underline{\mu}}$ 
$ {{M}}_{4\underline{m}}$, ${\Psi}^I_{4\underline{\mu}}$
${{M}}_{1\underline{m}}$  from \p{deSG}. 
Due to the identification \p{IIBstD9}, \p{IIBD3D9} only one set of independent fermionic variations $\d {\Theta}^{I\underline{\mu}} (x)$ is included into  
\p{deSG} and, thus,
the GFE action 
\p{Sint}  produces 
only one set of fermionic equations  
\begin{equation}\plabel{Theq}
J_8 \wedge \Psi_{2\underline{\mu}}^I  + J_6 \wedge \Psi_{4\underline{\mu}}^I  
=0.  
\end{equation}
However, as the only intersection of the worldsheet with the super-D-brane 
worldvolume is assumed to be just the boundary of the worldsheet 
${\cal M}^{1+3} \cap {\cal M}^{1+1}= \partial {\cal M}^{1+1}$ 
\p{inters}, 
Eq. \p{Theq} allows the statement  that
the pair of the fermionic equations \p{Theqs}
with indefinite source $\psi_{1\underline{\mu}}$ appears. 
Thus both methods of the description of the  interacting superbranes 
produce equivalent fermionic equations with an {\sl indefinite} source 
localized at the intersection. 
Actually some restrictions for the sources can be obtained 
using the explicit expressions for the l.h.s.-s of the fermionic equations 
\cite{BK}, 
but an ambiguity remains.

\bigskip 

Note that a similar ambiguity appears in the bosonic equations and, 
thus,  cannot be removed by passing to the pure bosonic limit.
Indeed, in accordance with 
\p{deSs} the bosonic equations 
\begin{equation}\plabel{Xeqs}
\hat{M}_{2\underline{m}} = 
j_1 \wedge \hat{\tilde{M}}_{1\underline{m}}  - 
j_1 \wedge P_{1\underline{m}}, \qquad 
\tilde{M}_{4\underline{m}}= -  j_3 \wedge P_{1\underline{m}} 
\end{equation}
involve an {\sl indefinite} Lagrange multiplier 1-form 
 $P_{1\underline{m}} $. 
The choice  $P_{1\underline{m}}=0$ corresponds to 
a sourceless equation for the host brane, 
which is the super-D3-brane in our case. 
Note that a definite source localized at the intersection 
is present in the supersymmetrized Born-Infeld equations, 
i.e. in the gauge field equations for the host brane.

 \section*{Conclusion and outlook}

In this note we propose two ways to obtain a 
supersymmetric action for interacting superbrane systems and 
present an explicit form of the actions 
for an open superstring ending on a super-D3-brane: \p{Sint01}, 
\p{LIIB}, \p{LD3}
 and \p{Sint}, \p{SD3+IIB}. 
They allow to obtain a manifestly supersymmetric 
(see Appendix A) set 
of equations of motion by straightforward variation.
One of the actions \p{Sint01} uses the Lagrange multiplier method 
to incorporate the necessary identification of the 
coordinate fields at the intersection, while the other 
(\p{Sint}, \p{SD3+IIB}) implies an identification of 
the Grassmann coordinate of intersecting branes with an image 
of the $D(=10)$--dimensional Goldstone fermion field. 
Thus such an action actually assumes the presence of an  
auxiliary or dynamical space-time filling brane and, hence,  
 can be called '(D-1)-brane dominance' model. 
An action for the space-time filling brane (in our case 
super-D9-brane) can be easily 
included in the action for an interacting low dimensional brane system 
like \p{Sint1}. 
On the other hand, it opens the  
possibility to include   
the supergravity into the coupled brane system: 
in a complete  action for a coupled brane system like \p{Sint1} 
the 
group-manifold action for $D$-dimensional supergravity 
may replace  the free action 
for the dynamical space-time filling brane.  

Inclusion of the (auxiliary or dynamical) space-time filling brane 
or of supergravity requires the use of the 
moving frame (Lorentz harmonic) actions \cite{BZ,bpstv,bsv,baku}
for low dimensional open branes and host branes. The reason is that  
their Lagrangian forms (in distinction to the ones of the 
standard actions) can be regarded as  pull--backs of some 
D-dimensional differential (p+1)--forms and, thus, the moving frame 
actions for free branes can be written easily in the form of integrals 
over a D--dimensional manifold by means of the current densities 
presented here (see \cite{bbs} for bosonic branes).   
Just the existence of the moving frame formulation may motivate the 
{\sl formal}  
 lifting of the Lagrangian forms of the standard actions 
to D dimensions and their use for the description of the interaction with 
space--time filling branes and/or supergravity.

 We studied the general structure of the equations of motion 
and found that for both approaches we arrive at an ambiguity in the source 
terms, which can be fixed only partially.  
Such an ambiguity actually appears 
as a result of the identification of the 
coordinate fields of the open brane and the host brane at the intersection. 
It  is inherent  
not only for the supersymmetric case, but for the pure bosonic limit of 
intersecting branes as well.

The explicit form of the equations of motion, the   
 analysis of their properties and the study of  $\kappa$--symmetry and 
supersymmetry issues for the action of interacting superbranes   
will be the subject of a forthcoming paper \cite{BK}.

\bigskip 
 
\section*{Acknowledgements} 

The authors are grateful to D. Sorokin and M. Tonin for interest in  
this work and many 
useful conversations and to 
R. Manvelyan and G. Mandal  for relevant discussions. 
One of the authors (I.B.) thanks the Austrian Science foundation for the 
support within the project {\bf M472-TPH}. 
A partial support from the INTAS Grant {\bf 96-308} and the 
Ukrainian GKNT grant {\bf 2.5.1/52} is acknowledged.  

\bigskip

\section*{Appendix A: On boundaries and supersymmetry}  
\setcounter{equation}{0}
\def\theequation{A.\arabic{equation}}

Our aim was to find the action which includes 
manifestly  
($N=2, D=10$) supersymmetric 
 'bulk' terms,  
 allows direct variations and, hence, leads to 
 equations of motion with manifestly supersymmetric l.h.s.-s. 
As it is well known, the presence of a boundary breaks the ($N=2$)  
supersymmetry of the classical action. 
For our system in the Lorentz harmonic formulation \p{Sint01}, \p{LIIB}, 
\p{LD3}, \p{Ea}, \p{harmvIIB}
the relevant boundary variation 
has the form 
\begin{equation}\plabel{varsusy}
(\d S)_{boundary} = \int_{\partial {\cal M}^{1+1}}
\left( {1 \over 2} {E}^{++} U_{\underline{m}}^{--}
- {1 \over 2} {E}^{--}  U_{\underline{m}}^{++}   
- E^b F_{ba} u^a_{\underline{m}}   
\right) i_\d \Pi^{\underline{m}}, 
\end{equation}
where we use the basis \p{varba} and put 
$ 
i_\d ({\cal F}-F)\equiv  \d A - \i_\d B_2 + E^b F_{ba} i_\d E^a=0$, 
which corresponds, in particular to the supersymmetry transformations 
of the gauge field $A$ (see \cite{c1}) which is chosen to make 
the super--D3--brane action supersymmetric. 
As mentioned in  Section 5, 
no boundary input 
with 
the variation $\d \Theta^I$  appears. 
This does not 
contradict the well--known fact that the presence of a 
worldsheet boundary breaks at least a 
half of the target space $N=2$ supersymmetry. 
Indeed,  for the supersymmetry 
transformations \p{susy1} the 
variation 
$i_\d \Pi^{\underline{m}}$ is nonvanishing and has the form 
\begin{equation}\plabel{Pisusy}
i_\d \Pi^{\underline{m}}= 2 \d X^{\underline{m}}=  2 \Theta^{I} \sigma^{\underline{m}} \e^I.  
\end{equation} 
Imposing the boundary conditions 
$\hat{\Theta}^{1\underline{\mu}}(\xi(\tau ))
 = \hat{\Theta}^{2\underline{\mu}} (\xi(\tau ))$ one arrives 
at the conservation of $N=1$ supersymmetry whose embedding into the 
type $IIB$ supersymmetry group is defined by 
$\e^{\underline{\mu}1}=-  \e^{\underline{\mu}2}$. 
Actually these conditions provide 
$i_\d \hat{\Pi}^{\underline{m}} (\xi(\tau ))=0$ and, as a consequence, 
the vanishing of 
the variation \p{varsusy}. 

The above consideration in the frame of the Lorentz harmonic approach 
results in an interesting observation that 
the supersymmetry breaking by boundary is related to the 
'classical reparametrization anomaly': indeed the variation \p{varsusy}, 
which produces the nonvanishing variation under $N=2$ supersymmetry 
transformation with \p{Pisusy}, contains only the 
variations $i_\d \Pi^{\underline{m}} U_{\underline{m}}^{\pm\pm}$ 
and $i_\d \Pi^{\underline{m}} u_{\underline{m}}^{a}$ , which 
correspond to reparametrization gauge symmetry 
of the free superstring and free super--D3--brane, respectively.

There exists a straightforward way to keep half of the 
rigid target space supersymmetry of the superstring--super-Dp-brane system 
by incorporation of the additional boundary  term 
$\int_{\partial{\cal M}^{1+1}} \phi_{1\underline{\mu}} 
\left(
\hat{\Theta}^{1\underline{\mu}}(\xi(\tau ))
 - \hat{\Theta}^{2\underline{\mu}} (\xi(\tau ))  
 \right)$ with a Grassmann Lagrange multiplier one form 
$\phi_{1\underline{\mu}}$. This  involves an additional arbitrariness 
in the first set of the fermionic equations \p{Theqs}, which now read 
$ \hat{\Psi}_{2\underline{\mu}}^I  = j_1 \wedge 
\left(\psi_{1\underline{\mu}}^I + (-1)^I \phi_{1\underline{\mu}} \right) 
$. 
 However, following \cite{Mourad,Sezgin25}, we accept in this paper the 
'soft' breaking of the supersymmetry by boundaries 
at the classical level (see \cite{HW,Mourad} for 
symmetry restoration by anomalies). 
We expect that the BPS states preserving  part of the target space 
supersymmetry 
will appear as particular solutions of the coupled superbrane equations
following from our actions. 

\section*{Appendix B} 
\setcounter{equation}{0}
\def\theequation{B.\arabic{equation}}

 In the search for a hypothetical  
generalization 
of our GFE action \p{Sint}, \p{SD3+IIB} 
the 
following completely 
supersymmetric counterpart of the current form \p{J80}  
can be useful
\begin{equation}\plabel{J8susy}
J_8 = \Pi_{\underline{m}\underline{n}}^{\wedge 8} \int_{{\cal M}^{1+1}}
\hat{V}_2^{\underline{m}\underline{n}} \delta^{10} (\hat{S}). 
\end{equation} 
In this equation 
\begin{equation}\plabel{Sinv}
\hat{S}^{\underline{m}} \equiv X^{\underline{m}} - 
\hat{X}^{\underline{m}} (\xi ) - i \Theta^I(X) \s ^{\underline{m}}
\hat{\Theta}^I  (\xi ) 
\end{equation}
is the supersymmetric invariant interval introduced in \cite{ZhT} for $D=4$. 
The measure $\hat{V}_2^{\underline{m}\underline{n}} $
can be constructed from supersymmetric invariant forms 
$\hat{\Pi}^{\underline{m}}$ and 
$$
\hat{d}\hat{S}^{\underline{m}}= 
-d\hat{X}^{\underline{m}} + i d\hat{\Theta}^1(\xi) \s^{\underline{m}}{\Theta}^1(X)
+ i d\hat{\Theta}^2(\xi) \s^{\underline{m}}{\Theta}^2(X)~~~: 
$$
\begin{equation}\plabel{V2}
\hat{V}_2^{\underline{m}\underline{n}1} = 
\hat{d}
\hat{S}^{\underline{m}} \wedge \hat{d}\hat{S}^{\underline{n}}, \qquad 
\hat{V}_2^{\underline{m}\underline{n}2} = 
\hat{\Pi}^{\underline{m}} \wedge \hat{\Pi}^{\underline{n}}, \qquad 
\hat{V}_2^{\underline{m}\underline{n}3} = 
\hat{\Pi}^{[\underline{m}} \wedge \hat{d}\hat{S}^{\underline{n}]}, 
\qquad  \ldots 
\end{equation}
The current form \p{J8susy} is invariant under 
the flat target space supersymmetry \p{susy1}
without the identification \p{IIBstD9},   
but assumes, nevertheless 
the presence of a space-time filling brane.   
However, an evident problem following this direction is the lack of a 
curved superspace generalization of the supersymmetric interval \p{Sinv}.



\end{document}